\documentclass[aps,pre,floatfix]{revtex4}
\usepackage{epsfig}
\usepackage{graphicx}
\usepackage{color}
\usepackage{floatflt}
\usepackage{lscape}
\usepackage{wrapfig}
\usepackage{setspace}
\begin{document}
\title{Boolean network model predicts cell cycle sequence of fission yeast}
\author{Maria I. Davidich}
\author{Stefan Bornholdt}
\affiliation{Institute for Theoretical Physics, University of
Bremen, D-28359 Bremen, Germany}
\bibliographystyle{apalike}
\begin{abstract}
A Boolean network model of the cell-cycle regulatory network
of fission yeast ({\em Schizosaccharomyces Pombe})
 is constructed solely on the basis of the known
biochemical interaction topology. Simulating the model in the
computer, faithfully reproduces the known sequence of
regulatory activity patterns along the cell cycle of the living cell.
Contrary to existing differential equation models, no parameters
enter the model except the structure of the regulatory circuitry.
The dynamical properties of the model indicate that the
biological dynamical sequence is robustly implemented in
the regulatory network, with the biological stationary state G1
corresponding to the dominant attractor in state space, and with
the biological regulatory sequence being a strongly attractive trajectory.
Comparing the fission yeast cell-cycle model to a similar model
of the corresponding network in {\sl S. cerevisiae}, a remarkable
difference in circuitry, as well as dynamics is observed. While
the latter operates in a strongly damped mode, driven
by external excitation, the {\sl S. pombe} network represents
an auto-excited system with external damping. 
\end{abstract}
\maketitle
\section*{Introduction}
Predicting the dynamics of complex molecular networks that control
living organisms is a central challenge of systems biology. While
cell-wide, or organism-wide, models of genetic and molecular
interactions appear well out of reach, predictive models of single
pathways and small modular molecular networks of living cells have
been studied with great success and are a matter of active research
\cite{Gunsalus2005,Hasty2001,Riel2006,Smolen2000}.

Given that the biochemical details of a chemical molecular network are
known, standard techniques are at hand for their computer
simulation. A method capturing molecular details is to use chemical
Monte-Carlo simulations \cite{Gillespie1976,Gillespie1977}, less
computationally costly and perhaps the most commonly used approach
to modeling biochemical pathways and networks are differential
equations which capture the underlying reaction kinetics in terms of
rates and concentrations \cite{Glass1973}. This method is highly
developed today and is broadly applied to predictive dynamical
modeling from single pathways to complex biochemical networks
\cite{Tyson2003}.

Such mathematical models contain detailed information about the time
evolution of the system which, in some circumstances, is more than
we are interested in. For many biological questions, knowledge of
the sequential pattern of states of the central control circuit of a
cell would be a sufficient answer, as, for example, in cell cycle
progression, cell commitment (e.g. to apoptosis), and in stem cell
control and differentiation. When we are interested in the path that
a cell takes, the exact time course of the control circuit dynamics
may not be needed, however, its modeling takes most effort and often
one needs to know large numbers of biochemical parameters that are 
not easily obtained \cite{Sveiczer2000, Tyson2001a}. 
 
Indeed, recent research indicates that some molecular control 
networks are so robustly designed that timing is not a critical 
factor \cite{Braunewell2006}. Vice versa, as a working hypothesis, 
this observation bears the chance for vastly simplified dynamical 
models for molecular networks, as soon as one drops the requirement
for accurate reproduction of timing by the model, just asking for
the sequence of dynamical patterns of the network. Recent studies
demonstrate, that such more simplified models indeed can reproduce
the sequence of states in biological systems. For example, a class
of discrete dynamical systems with binary states, mathematically
similar to models used in artificial neural networks, has recently
proven to predict specific sequence patterns of expressed genes as
observed in living cells \cite{Albert2003,Li2004}.

Such models are in the mathematical tradition of random Boolean
networks which, for decades, served as a simplistic analogy for how
gene regulation networks could in principle work \cite{Kaufmann}. In
these historical studies, dynamical properties of random networks of
discrete dynamical elements were studied to derive possible
properties of (the then hardly known) regulatory circuits \cite{Kauffman1993}. 
In the new approach outlined above, however, similar mathematical elements 
now serve to simulate one specific known biological control network. 
From a different perspective, they can be viewed as a further simplification 
of the differential equation approach \cite{Bornholdt2005}. Recent
application of this model class to modeling real biological genetic
circuits show that they can predict expression pattern sequences
with much less input (e.g. parameters) to the model as the classical
differential equations approach. Examples are models of the genetic
network of {\em A.\ thaliana}
\cite{Espinosa-Soto2004a,Mendoza1999,Thum2003}, the
cell-cycle networks of {\em S. cerevisiae} \cite{Li2004} and of the
mammalian cell-cycle \cite{Faure2006}, as well as the segment
polarity gene network in {\em D.\ melanogaster} \cite{Albert2003,
Sanchez2001}.

For example, the model by Albert and Othmer \cite{Albert2003} of the
segment polarity gene network in {\em D.\ melanogaster}, as well as
the model by Li et al.\ \cite{Li2004} of the {\em S.\ cerevisiae} cell-cycle
control network, yield accurate predictions of sequential expression
patterns, previously not obtained from such a simple model class.
In these models, the dynamics can be viewed in terms of flow in state
space of possible states of the network, converging towards so-called 
attractors, or fixed points, which here correspond to specific biological 
states. These attractors and their basins of attraction in state space 
mainly depend on the circuitry of the network, and their analysis yields 
further information about the robustness of the dynamics against errors
or mutations.

How generic is this approach? In this article we address the question
whether the approach of discrete dynamical network models
is a more general method, namely whether constructing predictive
dynamical models for gene regulation from Boolean networks is a
straightforward procedure that generalizes to other organisms.
We choose the fission yeast ({\em Schizosaccharomyces Pombe})
cell-cycle as an example system that on the one hand is well understood
in terms of conventional differential equation models, but on the other hand 
is markedly different from the above examples, as {\em S.\ cerevisiae}.
{\em S.\ Pombe} has been sequenced in 1999 and has been used as a 
model organism only relatively recently \cite{Forsburg}.
Models exist \cite{Novak1997,Novak2001} that mathematically
model the fission yeast cell-cycle with a common ODE (ordinary
differential equation) approach. These are based on a set of differential
equations for the biochemical concentrations that take part in the network
and their change in time (and space). This approach allows to
predict the dynamics of the fission yeast cell-cycle for the wild-type
and some known mutant cells \cite{Tyson2001a, Tyson2002b}.

We will in the following construct a discrete dynamical model for
the fission yeast cell cycle network. An interesting question will be,
how far we will get without considering parameters, as kinetic constants 
etc., that are a key ingredient of the existing models. We will base our model on
the circuitry of the known biochemical network, only. Let us in the next
section briefly review the fission yeast cell cycle network, then define
our discrete dynamical model in the subsequent section. This is followed
by a section reporting our results, and then we will compare our findings
with a similar model of the budding yeast ({\em S.\ cerevisiae}) network
and conclude with a discussion.

\section*{The fission yeast cell cycle network}
Let us briefly review the regulatory processes that control the cell
cycle in {\em Saccharomyces Pombe}. The full process of one cell
division consists of four stages, named G1--S--G2--M. At the first
stage (G1), the cell grows and, under specific conditions, commits
to division. At the second stage (S), DNA is synthesized and
chromosomes are replicated. This is followed by a "gap" stage G2.
The final stage (M) corresponds to mitosis, in which chromosomes are
separated and the cell divides itself. Eventually, after the M stage, 
the cell enters G1 again, thereby completing one cycle.

The biochemical reactions that form the network that controls the
fission yeast cell-cycle have been studied in detail over the last
years \cite{Novak2001,Buck2003,Correabordes1995,Jaspersen1999,
Lundgren1991,Martin-Castellanos1996,Russel1987,
Visintin1998,Yamaguchi2000}. The major role is played by a
cyclin-dependent protein kinase complex Cdc2/Cdc13 with a
protein Tyr-15, a residue of Cdc2. Tyr-15 acts as a label for high
Cdc2/Cdc13 concentration. It is inactive during the G2 phase, when
Cdc2/Cdc13 is phosphorylated, and becomes active during the G2--M
transition \cite{Novak2001,Tyson2002b}. The other members that
participate in the cell-cycle control can be attributed to two
different classes. The first class consists of positive regulators
of the kinase Cdc2/Cdc13: "Start kinase" (SK), a group of Cdk/cyclin
complexes (Cdc2 with Cig1, Cig2 and Puc1 cyclins), and the
phosphatase Cdc25. A second class is composed of the antagonists of
the complex Cdc2/Cdc13: Slp1, Rum1, Ste9, and the phosphatase PP
\cite{Sveiczer2000}.

We give a full compilation of the network of key-regulators of the
fission yeast cell cycle network in Table 1, corresponding to our
current knowledge as given in 
\cite{Sveiczer2000,Novak2001,Tyson2002b}. 
\begin{table}
\footnotesize
\begin{center}
\begin{tabular}{|p{3cm}|p{3cm}|p{3cm}|p{3cm}|}
\hline Parent node & Daughter node &   Rule of activation
(comments) &   Rule of inhibition (comments) \\ \hline Start node &
Starter Kinases (SK): Cdc2/Cig1, Cdc2/Cig2, Cdc2/Puc1& Starter
kinases, take part in first transition G1/S, +1 \cite{Sveiczer2000}. & \hspace{0cm} \\
\hline SK & Ste9, Rum1 & & Phosphorylate, thereby inactivate, -1 \cite{Sveiczer2000, Tyson2002b}\\
\hline Cdc2/Cdc13 & Cdc25 & Cdc25 is phosphorylated thereby activated, +1 \cite{Sveiczer2000}.&  \\
\hline Wee1, Mik1 & Tyr15 & &     Phosphorylate, inactivating, -1
\cite{Sveiczer2000}
\\ \hline Rum1 &   Cdc2/Cdc13
& & Binds and inhibits activity, -1 Cdc2/Cdc13 \cite{Sveiczer2000}. \\
\hline Cdc2/Cdc13 & Rum1 & & Phosphorylates and thereby targets Rum1
for degradation. -1 \cite{Sveiczer2000, Tyson2002b}
 \\ \hline Ste9 &   Cdc2/Cdc13 & &     Labels Cdc13 for
degradation \cite{Tyson2002b,Sveiczer2000}, -1. \\ \hline Tyr15,
Cdc2/Cdc13 & Slp1 & Highly activated Cdc2/Cdc13 activates Slp1,
Tyr15 has to be active, too \cite{Novak2001, Sveiczer2000}+1. & \hspace{0cm}
\\  \hline Slp1  &  Cdc2/Cdc13 & & Promotes degradation of
Cdc13, thereby the activity of Cdc2/Cdc13 drops -1
\cite{Sveiczer2000}
\\ \hline Slp1 &   PP  &  Activates, +1 \cite{Sveiczer2000} &
\hspace{0cm}
\\ \hline PP(Unknown phosphatase)& Ste9, Rum1, Wee1, Mik1 & Activates Rum1,
Ste9, and the tyrosine-modifying enzymes (Wee1, Mik1, \cite{Sveiczer2000}, +1 & \\
\hline Cdc25  & Tyr15 & Cdc25 reverses phosphorylation of Cdc2,
thereby Tyr15 becomes active, +1 \cite{Sveiczer2000, Novak2001} & \hspace{0cm} \\
\hline Wee1,
Mik1 & Tyr15 & Phosphorylates, thereby inactivating, -1 \cite{Novak2001} & \hspace{0cm}\\
\hline PP  &  Cdc25 & & inhibits, \cite{Sveiczer2000} -1 \hspace{0cm}
\\
\hline
\end{tabular}
\caption {The rules of interaction of the main elements involved in
the fission yeast cell cycle regulation.}
\end{center}
\end{table}
Also our translation 
into an interaction graph with activating and inhibiting links is 
given in the table, which is the starting point for our discrete
dynamical network simulation of this network. Let us in the next
section define the discrete dynamics that we will simulate on this
graph.

\section*{A discrete dynamical model of the cell cycle network}
We assume proteins to be the nodes of the network and assign a
binary value $S_i(t) \in \{0,1\}$ to each node $i$, denoting whether
the protein is present or not (due to different possible biochemical
mechanisms, as, e.g., gene expression of a corresponding protein, or
fast biochemical reactions as phosphorylization). The interactions
between the nodes, as compiled in Table 1, are denoted as links, or
arrows, see Figure 1. 
\begin{figure}
\begin{center}
\includegraphics [width=12cm]{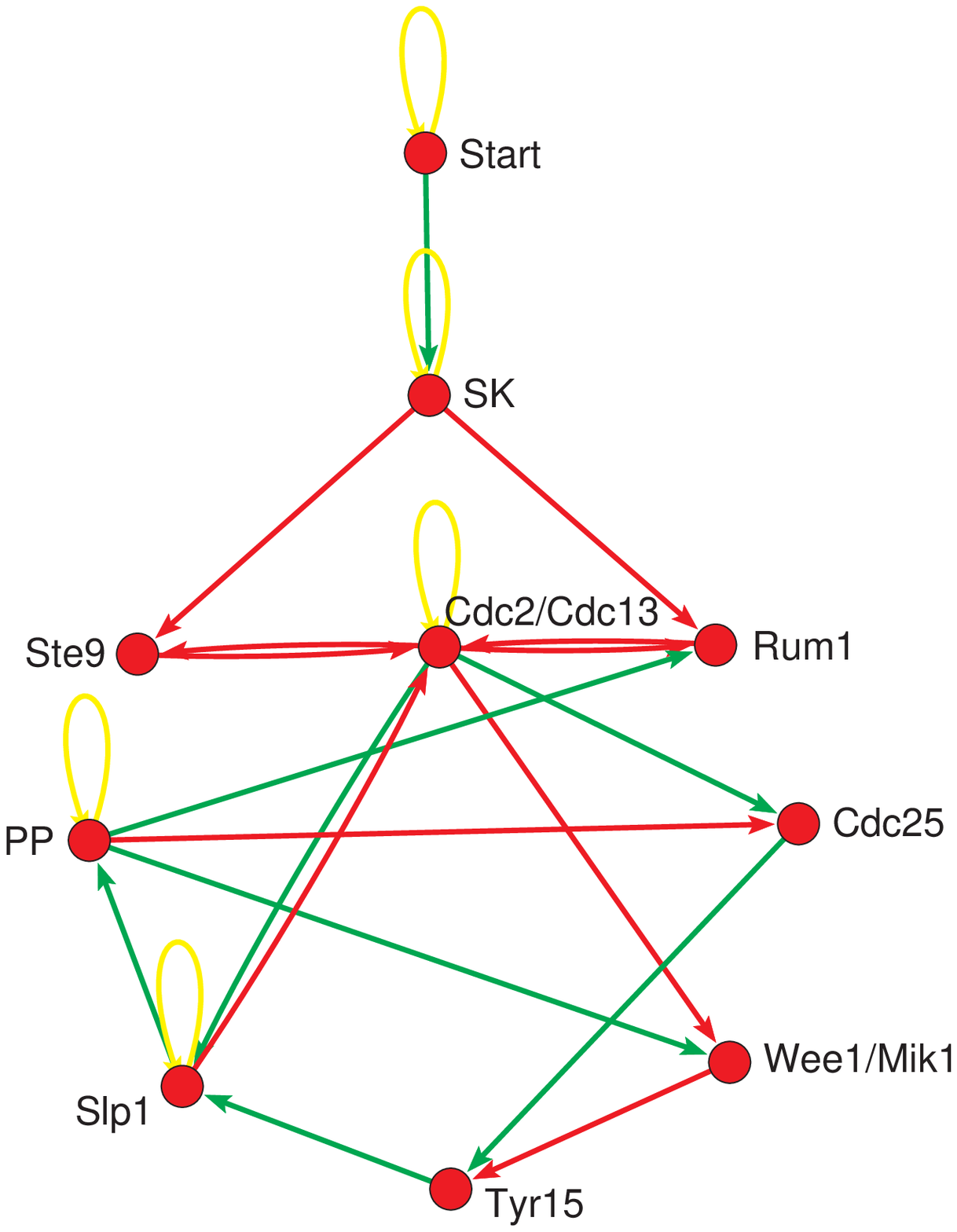}
\caption {Network model of the fission yeast cell-cycle regulation.}
\end{center}
\label{cell-cycle}
\end{figure}
We do not quantify any interaction strength,
except whether a link is present or not, and whether it is
activating or inhibiting. Again, different biochemical mechanisms
are subsumed under this simplified picture, as, e.g.,
transcriptional regulation, or faster enzymatic interactions. The
dynamics of the nodes are updated (in parallel) in discrete time
steps according to the following rule:
\begin{equation}
\it
 S_i(t+1)= \left\{
\begin{array}{cc}
1, & \sum_j a_{ij}S_j(t)>0, \\
0, & \sum_j  a_{ij}S_j(t)<0, \\
S_i(t) & \sum_j  a_{ij}S_j(t)=0,
\end{array}
\right. \label{common_rule}
\end{equation}
where $a_{ij}=1$ for an activating interaction (green link) from node 
${j}$ to node ${i}$, and $a_{ij}=-1$ for an inhibiting (red) link from node 
${j}$ to node ${i}$, and $a_{ij}=0$ for no interaction at all. 
This definition follows closely the approach in \cite{Li2004}.
The dramatic simplification steps in constructing this model consist
in not differentiating between absolute values of interaction
strengths on the one hand, and not distinguishing between the
different time scales of the biochemical interactions involved on
the other. This corresponds to dropping all biochemical parameter
values, time constants as well as binding constants, from the
differential equation models. As we will see below, dynamical models
on networks can be built to be insensitive to these parameters,
provided that the interaction topology has certain properties.

Two of the ten proteins included in the model exhibit a slightly different 
activation behavior, which we account for by the two following rules 
(which alternatively could be incorporated into the above equations 
as a non-zero activation threshold). 
Slp1 is only activated by a highly active complex Cdc2/Cdc13, 
which corresponds to both active Cdc2/Cdc13 as well as active Tyr15, 
since Tyr15 labels the level of activity of Cdc2. 
This mechanism acts as a barrier for entering mitosis. 
The second special rule is to add "self-activation" (corresponding to 
adding a negative activation threshold) to the node Cdc2/Cdc13, 
as it is otherwise not positively regulated. 

We also follow \cite{Li2004} by adding "self-degradation" (yellow
loops) to those nodes that are not negatively regulated by others,
representing the continuous degradation of proteins in the cell,
which corresponds to $a_{ii}=-1$.

Nodes, that have the same function as, for example, Wee1/Mik1 and SK
(Cdc2/Cig1, Cdc2/Cig2, Cdc2/Puc1) are joined together in a single node
(see Figure 1), as it does not make a difference in the specific mathematical
model dynamics considered here.

Finally let us define the initial condition of the model at the start of the
simulation, which is chosen to correspond to the biological start condition, 
i.e.\ all nodes being in the OFF (inactive) state, except for the proteins
Start, Ste9, Rum1, and Wee1/Mik1 \cite{Tyson2002b}.

\section*{Results}
\subsection*{Simulation of the fission yeast cell cycle}
Let us first consider the time evolution of the proteins of the
dynamical model described above. Let us run the cell-cycle model 
by exciting the G1 stationary state with the cell size signal 
("Start" node). This initiates a sequence of network activation 
states ("expression patterns") that, eventually, return to the G1 
stationary state. The temporal
evolution of the protein states is presented in Table 2, 
\begin{table}
\footnotesize
\begin{tabular}{|p{0,7cm}|p{0,7cm}|p{0,3cm}|p{0,85cm}|p{0,45cm}|p{0,65cm}|p{0,45cm}|p{0,7cm}|p{0,55cm}|p{0,7cm}|p{0,5cm}|p{0,9cm}|p{3,1cm}|}
\hline  Time Step& Start & SK &Cdc2 /Cdc13&Ste9&Rum1&Slp1&Tyr15&Wee1 Mik1& Cdc25&PP & Phase &  comments\\
\hline  1 &   1 &   0 & 0 & 1 & 1 & 0 & 0 & 1 & 0 & 0 & START &
Cdc2/Cdc13 dimers are inhibited, antagonists are active.\\
\hline  2 & 0 & 1 & 0 & 1 &   0 & 0 &   0 &   1 & 0 &   0 &   G1 &
SK are becoming active \\
\hline  3 & 0 & 0 & 0 & 0 & 0 & 0 & 0 &   1 & 0 &  0 & G1/S &
When Cdc2/Cdc13 and SK dimers switch off Rum1 and Ste9/APC, the
cell passes 'Start' and DNA replication takes place, 
Cdc2/Cdc13 starts to accumulate \\
\hline   4 &   0 & 0 & 1 & 0 & 0 & 0 & 0 & 1 & 0 & 0 & G2 & Activity
of Cdc2/Cdc13 achieves moderate level, which is enough for entering
G2 phase but not mitosis, since Wee1/Mik1 inhibits the
residue of Cdc2--Tyr15 \\
\hline  5 &   0 & 0 & 1 & 0 & 0 & 0 & 0 &  0 &   1 & 0 & G2 & 
moderate activity Cdc2/Cdc13 activates Cdc25\\
\hline  6 &   0 & 0 & 1 & 0 & 0 & 0 & 1 & 0 & 1 &   0 & G2/M & Cdc25
reverses phosphorylation, removing the inhibiting phosphate group
and activating Tyr15 \\
\hline   7  & 0 &  0 &   1 & 0  & 0  & 1 & 1 & 0 & 1 & 0 & G2/M &
Cdc2/Cdc13 reaches high activity level sufficient to activate 
Slp1/APC (Cdc2/Cdc13 and Tyr15 are both active) and cell enters
mitosis\\ 
\hline  8 & 0 & 0 &  0 &  0 & 0 & 1 & 1 & 0 & 1 & 1& M & Slp1
degrades Cdc13 and activates unknown phosphase \\
\hline  9 &   0 & 0 & 0 & 1 & 1 & 0 & 1 & 1 & 0 & 1 & M &
Antagonists of Cdc2/Cdc13 are reset \\
\hline  10 &  0 & 0 & 0 & 1 & 1 & 0 & 0 & 1 & 0 & 0 & G1 & 
Cdc13 is degraded, Cdc2 thereby downregulated, 
cell reaches G1 stationary state \\
\hline
\end{tabular}
\caption{Temporal evolution of protein states in the cell cycle network.}
\end{table}
where one
observes a sequence of states which exactly matches the
corresponding biological expression pattern along the cell-cycle,
from the excited G1 state (START) through S and G2 to the M phase
and finally back to the stationary G1 state. It corresponds to the 
biological time sequence of the protein states in the cell-cycle 
control network. This is a remarkable observation as it is unlikely 
to occur by chance due to the size of the state space.

In the next step we run the model starting from each one of the 
$2^{10}=1024$ possible initial states. We find that each initial state 
flows into one of 15 stationary states (fixed points). The largest attractor 
belongs to a fixed point attracting $77\%$ of all network states. Our 
first observation is that this fixed point exactly coincides with the biological 
G1 stationary state (see Table 3) of the cell. 
\begin{table}
\footnotesize
\begin{center}
\begin{tabular}{|r|r|c|c|c|c|c|c|c|c|c|c|}
\hline  
Attractor&Basin size&Start&SK&Cdc2/Cdc13&Ste9&Rum1&Slp1&Tyr15& Wee1/Mik1& Cdc25  & PP\\
\hline 1 & 788 & 0&   0&   0&   1&   1&   0&   1&   0&   0& 0 \\
\hline 2  & 136 & 0 & 0 & 0 & 0 & 1 & 0 & 0 &   1 &   0 & 0\\
\hline 3 &  33&   0& 0 & 0 & 1 & 0 & 0 & 1 & 0& 1 & 0\\
\hline 4 & 28 & 0 & 0 & 0 & 1 & 0 & 0 &  0 & 1 & 0 & 0\\
\hline 5 & 11 & 0 & 0 &   0&   1 & 0& 0& 0& 1& 1 & 0\\
\hline 6 & 8 & 0 & 0 & 0 & 1 & 0 & 0 & 1 & 1 & 1& 0 \\
\hline 7 & 6& 0 & 0 & 0 & 1 & 1 & 0 & 1 & 0 & 1 & 0\\
\hline 8 & 4 & 0 & 0 & 1 & 0 & 0 & 0 & 1 & 0 & 0 & 0\\
\hline 9  & 3 & 0 & 0 & 0 & 0 & 1 & 0 & 0 & 0 & 0 & 0\\
\hline 10 & 2 & 0 & 0 & 0 & 1 & 1 & 0 & 1 & 1 & 1 & 0\\
\hline 11 & 1 & 0 & 0 & 0 & 1 & 0 & 0 & 0 & 0 & 0 & 0 \\
\hline 12 & 1 & 0 & 0 & 0 & 1 & 0 & 1 & 0 & 0 & 0 & 0 \\
\hline 13 & 1 & 0 & 0 & 0 & 1 & 0 & 0 & 1 & 0& 0 & 0 \\
\hline 14 & 1 & 0 & 0  & 0 & 1 & 1 & 0 & 0 & 0 & 0 & 0\\
\hline 15 & 1 & 0 & 0 & 0 & 1 & 1 & 0 & 1 & 0 & 0 & 0\\
\hline
\end{tabular}
\caption{All attractors (fixed points) of the dynamics of the network model 
for the fission yeast cell cycle regulation.} 
\end{center}
\end{table}
Thus, the biological
target state is the dominant attractor of the network dynamics. As
soon as the system reaches this state with the specific
corresponding combination of active and inactive proteins, it stays
there, and is likely to do so even in the presence of perturbations.

\begin{figure}
\begin{center}
\includegraphics [width=16cm]{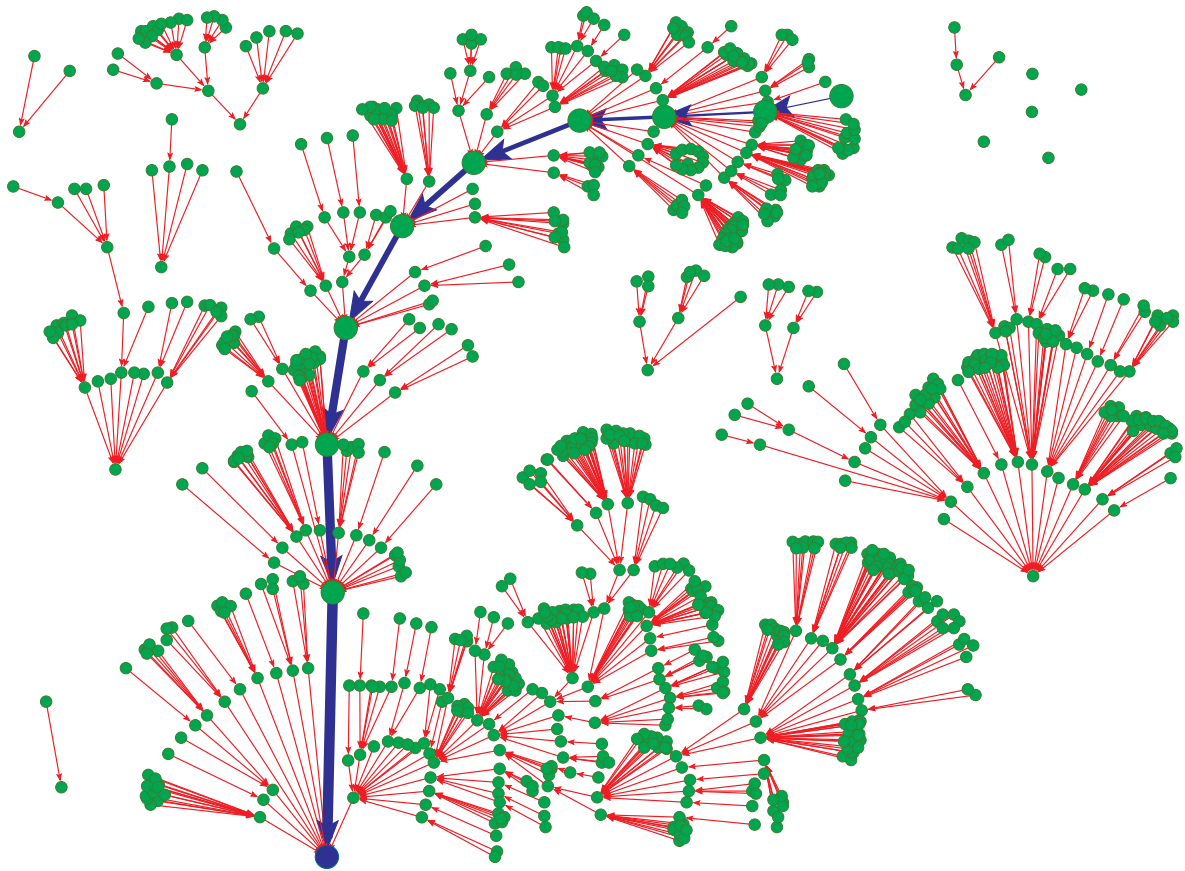}
\end{center}
\caption {State space of the 1024 possible network states (green circles) 
and their dynamical trajectories, all converging towards fixed point 
attractors. Each circle corresponds to one specific network state 
with each of the ten proteins being in one specific activation state
(active/inactive). 
The largest attractor tree corresponds to all network states flowing 
to the G1 fixed point (blue node). Arrows between the network 
states indicate the direction of the dynamical flow from one network 
state to its subsequent state. The fission yeast cell-cycle sequence 
is shown with blue arrows.}
\end{figure}
A further observation is best depicted by Figure 2, showing the
dynamical flow of the network states, and how it converges towards the 
biological fixed point. In this figure, the dynamical trajectories in the state 
space starting from all 1024 possible initial states of the network are shown.
Each network state is represented by a dot, with the arrows between
them indicating the dynamical transition from one state to its temporally
subsequent state. At the root of the largest attractor (tree) the G1 state
is found and the blue arrows show the biological time sequence that
leads to it. This attractor tree consists of $77\%$ of all network states.

We further performed a robustness test by reversing the state of a single, 
randomly chosen node while the network proceeds through the biological 
sequence. This deviation from the biological pathway by the activity state 
of one single protein at one randomly chosen step of the cycle, the 
system returns to the fixed point G1 in 90 out of 100 possible cases.
Thus we observe an additional robustness in the fission yeast cell-cycle
network, meaning that there is an increased probability to stay in the
attractor basin of the biological fixed point when perturbing states along
the biological trajectory.

An immediate question about the specific network structure considered
here is whether the architecture of the network has special properties
as, for example, traces of being optimized by biological evolution.
We  compare the network dynamics to the null model of random networks
with the same number of inhibiting and activating links, self-degrading
and self-activating nodes and the same activation thresholds.
Indeed one finds that the corresponding random networks typically have
 smaller attractors. The mean size of the biggest attractors is about $38\%$
 of all initial states (averaged over 1000 random networks). This may
 indicate that attractor basin size of the biological attractor is optimized,
 possibly in order to provide additional dynamical robustness.

\subsection*{Comparison with {\em S.\ cerevisiae}}
The two yeasts, {\em S.\ cerevisiae} and {\em S.\ pombe}, are remarkably
different cells and a comparison may provide insights relevant
for the understanding of higher eukaryotic organisms. As we now
have discrete dynamical models for the cell cycle network of both
of them at hand (this work, as well as \cite{Li2004}), let us discuss
how they compare.

As these two organisms are closely related genetically, one might
expect a large overlap also in the biochemical control machinery. On
the other hand, the biology of the two is markedly different, so
there have to be some differences on the biochemical level as well.
As an overview, the second model is shown in Figure 3.
\begin{figure}
\begin{center}
\includegraphics[width=12cm]{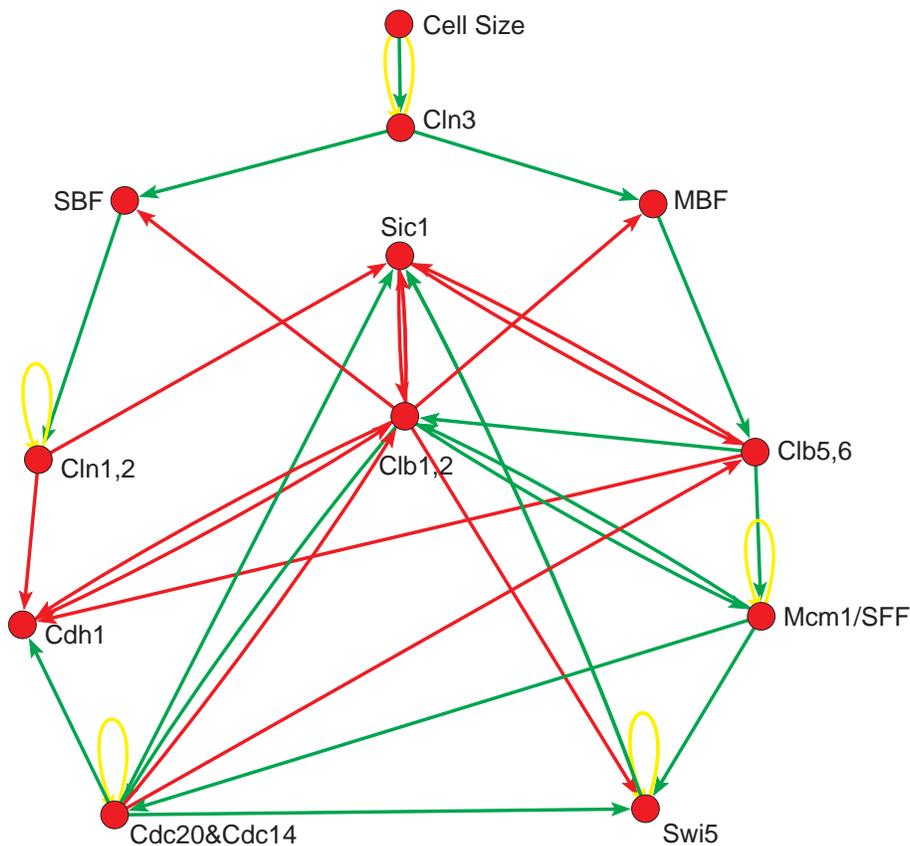}
\end{center}
\caption {Budding yeast cell cycle network model of \cite{Li2004}, 
for comparison with our model of fission yeast. This network relies 
more on transcriptional regulation than the fission yeast network 
(note that some homologues corresponding to the latter do not
have to be included here). Note also the difference in circuitry.} 
\label{cell-cycle}
\end{figure}

There are a number of closely related genes (see Table 4) between
the two yeasts \cite{Tyson2001a} which, however, can have vastly 
differing functions \cite{Forsburg}. 
\begin{table}
\begin{center}
\begin{tabular}{|p{2,1cm}|p{1,5cm}|p{1,5cm}|p{1,5cm}|p{1,5cm}|p{1,5cm}|}
\hline  Fission yeast   & Rum1 &    Ste9 &    Slp1 &    Cdc2 & Cdc13\\
\hline Budding yeast  & Sic1 &    Cdh11 &   Cdc20 &   Cdc28 &
Clb1-6\\
\hline
\end{tabular}
\caption{Homologue proteins related to the cell cycle networks of fission yeast 
and budding yeast}
\end{center}
\end{table}
In fission yeast, for example, 
phosphatase Cdc25 is required for the G2--M transition, while in the 
model of budding yeast \cite{Li2004} the corresponding homologue 
Mih1 is insignificant. The reason is that in the fission yeast cell cycle,
Cdc25 removes an inhibitory phosphate group from the residue Tyr-15
of Cdc2, which is important for the right timing of the G2--M transition. 
In contrast, the tyrosine residue in {\em S.\ cerevisiae} Cdc28 kinase 
(fission yeast: Cdc2) is not as critical and usually not phosphorylated. 
Therefore, for a model of fission yeast, Cdc25 is essential, whereas 
the homologue Mih1 in budding yeast is not \cite{Li2004}. One other 
example is the role of the protein Cdc13. In fission yeast it acts in a 
complex with Cdc2, while in the budding yeast model its functionality 
is represented by two complexes Clb1,2/Cdc28 and Clb5,6/Cdc28, 
which exhibit some differences in interactions, as well as in timing.

Despite of  the differences in many details, the general logic of both 
yeast cell cycles is surprisingly similar and exhibits a number of
"structural homologues". For example both exhibit a negative
feedback loop similar in role: Clb1,2/Cdc28 activates Cdc20 which
inhibits Clb1,2/Cdc28 (fission yeast: Cdc2/Cdc13, Tyr15 activate
Slp1,which inhibits Cdc2/Cdc13).

The most interesting comparison is in our view on the level of
the global network dynamics. From this point of view, the
{\em S.\ cerevisiae} network is a strongly damped system, driven
by external excitation. External signals are entering the network,
triggering signal cascades in the network that induce the subsequent
phases. In contrast, the network of {\em S.\ pombe} corresponds
to an auto-excited system (there are two nodes with self-excitation
- Cdc2/Cdc13 and Wee1/Mik1) with additional damping.
Here, an external signal works as a trigger mechanism that
counteracts internal damping, causing the auto-excitation to spread
its activity in the system.

While these differences in the "mechanics" of the signalling
networks are considerable, the overall dynamics is surprisingly
similar. The state space picture is quite similar in both cases: one
observes only a small number of attractors and just one big global
attractor (with $86\%$ resp.~$77\%$ of all initial states) which for
both organisms corresponds to the stationary G1 state.

Finally, a most prominent difference between the two yeast networks
is their choice in biochemical machinery: {\em S.\ cerevisiae} relies
more on transcriptional factors while {\em S.\ pombe} mostly relies
on post-translational regulation \cite{Simanis2003}. From the
methodological point of view, we note that for this reason we were
surprised to find our model for  the {\em S. pombe} cell cycle
network so robust against neglecting the vastly different time
scales of interactions, which we expected to be the major difficulty
in constructing a discrete dynamical model for {\em S. pombe} as 
compared to {\em S.\ cerevisiae}.

\section*{Discussion}
We have constructed a Boolean model for the biochemical network that
controls the cell cycle progression in fission yeast  {\em S.\
pombe}, and found a number of interesting results. The dynamics of
this network reproduces the time sequence of expression patterns
along the biological cell cycle, solely on the basis of the
connectivity graph of the network, neglecting all biochemical
kinetic parameters. The dynamics of the network is characterized by
a dominant attractor in the space of all possible states, with an
attractor basin that attracts most of all states. The network
dynamics are robust against perturbation of the biological
expression pattern.

The results obtained from our model are in accordance with the existing
ODE model of fission yeast \cite{Tyson2001a}. Let us discuss the
differences between these two approaches. The  {\em S.\ pombe} ODE system
\cite{Tyson2001a} has several steady state solutions. One can identify
every such solution with the corresponding physiological stage.
The growth of cell size brings the cell from one phase to another via
a series of bifurcations. At the same time, other variables indicate the
degree of activity of various components of the cell regulatory nodes.
One observes \cite{Tyson2002b} that the typical curves depicting this
activity have almost rectangular shape. This motivates our choice of
binary valued function to approximate protein concentrations in time.
Further, the ODE-based model makes use of continuous system
parameters, which we omit and replace by their signs, only. As
a result, the ODE bifurcation curve then corresponds to the Boolean
biological path. The main advantage of our Boolean model is that we
were able to drop 47 kinetic constants that were necessary in the ODE
approach and, while doing so, still reproduce the biological activation 
pattern of the system.

This fact and our further observations point at built-in dynamical robustness
of the network, which may provide a further mechanism for organisms to ensure
functional robustness \cite{Alon1999}. In return, our study indicates that
the regulatory robustness of biological chemical networks may allow for
"robust" modeling approaches: Our paradigm here is nothing but assuming
that biochemical networks are functioning in a parameter-insensitive way ---
which motvated us to eliminate all tunable parameters from the model. That our 
model reproduces the biological sequence instantly without any further parameter 
tuning, confirms our assumption {\em a posteriori}.  We therefore encourage 
further modeling experiments with the here presented quite minimalistic approach, 
as it may prove a quick approach to predicting biologically relevant dynamical 
features of genetic and protein networks in the living cell.

\end{document}